\documentclass[aps,prd,twocolumn,superscriptaddress]{revtex4}
\usepackage{epsfig,epsf}
\usepackage{amsmath}
\usepackage{amsthm}
\usepackage{amsfonts}
\usepackage{amssymb}
\usepackage{dsfont}
\usepackage{multirow}
\usepackage{appendix}
\usepackage{slashed}
\usepackage[active]{srcltx}
\usepackage{psfrag}

\begin{document}

\title{Exploring the resonances $X(4140)$ and $X(4274)$ through their decay
channels }
\date{\today}
\author{S.~S.~Agaev}
\affiliation{Institute for Physical Problems, Baku State University, Az--1148 Baku,
Azerbaijan}
\author{K.~Azizi}
\affiliation{Department of Physics, Do\v{g}u\c{s} University, Acibadem-Kadik\"{o}y, 34722
Istanbul, Turkey}
\author{H.~Sundu}
\affiliation{Department of Physics, Kocaeli University, 41380 Izmit, Turkey}

\begin{abstract}
Investigation of the resonances $X(4140)$ and $X(4274)$, which were recently
confirmed by the LHCb Collaboration \cite{Aaij:2016iza}, is carried out by
treating them as the color triplet and sextet $[cs][\bar{c}\bar {s}]$
diquark-antidiquark states with the spin-parity $J^{P}=1^{+}$, respectively.
We calculate the masses and meson-current couplings of these tetraquarks in
the context of QCD two-point sum rule method by taking into account the
quark, gluon and mixed vacuum condensates up to eight dimensions. We also
study the vertices $X(4140)J/\psi\phi$ and $X(4274)J/\psi\phi$, and evaluate
corresponding strong couplings $g_{X(4140)J/\psi\phi}$ and $%
g_{X(4274)J/\psi\phi}$ by means of QCD light-cone sum rule method, and a
technique of the soft-meson approximation. In turn, these  couplings
contain a required information to determine the width of the
$X(4140)\to J/\psi\phi$ and $X(4274) \to J/\psi\phi$ decay channels.
We compare our results for the masses and decay widths of the $X(4140)$
and $X(4274)$ resonances with the LHCb data, and alternative theoretical
predictions.
\end{abstract}

\maketitle

\section{Introduction}

Recently the LHCb Collaboration presented results of analysis of the
exclusive decays $B^{+}\rightarrow J/\psi \phi K^{+}$, and confirmed
existence of the resonances $X(4140)$ and $X(4274)$ in the $J/\psi \phi $
invariant mass distribution \cite{Aaij:2016iza}. It also reported on
observation of the heavy resonances $X(4500)$ and $X(4700)$ in the same $%
J/\psi \phi $ channel. The measured masses and decay widths of these
resonances (hereafter $X(4140)\Rightarrow X_{1},\ X(4274) \Rightarrow
X_{2},\ X(4500)\Rightarrow X_{3}$ and $X(4700)\Rightarrow X_{4}$,
respectively ) read
\begin{eqnarray}
&&X_{1}:M=4146\pm 4.5_{-2.8}^{+4.6}\ \mathrm{MeV},\ \Gamma =83\pm
21_{-14}^{+21}\ \mathrm{MeV},  \notag \\
&&X_{2}:M=4273\pm 8.3_{-3.6}^{+17.2}\ \mathrm{MeV},\ \Gamma =56\pm
11_{-11}^{+8}\ \mathrm{MeV},  \notag \\
&&X_{3}:M=4506\pm 11_{-15}^{+12}\ \mathrm{MeV},\ \Gamma =92\pm
21_{-20}^{+21}\ \mathrm{MeV},  \notag \\
&&X_{4}:M=4704\pm 10_{-24}^{+14}\ \mathrm{MeV},\ \Gamma =120\pm
31_{-33}^{+42}\ \mathrm{MeV}.  \notag \\
&&{}
\end{eqnarray}%
The LHCb determined the spin-parities of these resonances, as well. It
turned out, that $X_{1}$ and $X_{2}$ are axial-vector states with $%
J^{PC}=1^{++}$, whereas the quantum numbers of $X_{3}$ and $X_{4}$ are $%
J^{PC}=0^{++}$.

The resonances $X_{1}$ and $X_{2}$ are old members of the XYZ family of
exotic states: They were  observed by the CDF Collaboration \cite%
{Aaltonen:2009tz} in the decay processes $B^{\pm }\rightarrow J/\psi \phi
K^{\pm }$, and later confirmed by CMS \cite{Chatrchyan:2013dma} and D0
collaborations \cite{Abazov:2013xda}, respectively. The states $X_{3}$ and $%
X_{4}$ are heavier than $X_{1},\,X_{2}$, and were found for the first time.
All of the $X$ resonances may belong to a class of the hidden-charm exotic
states. From production mechanisms and decay channels, it is clear that as
tetraquark candidates they should contain strange quark-antiquark pair $s%
\bar{s}$. In other words, the quark content of the $X$ states is $c\bar{c}s%
\bar{s}$.

The unconventional hadrons, such as glueballs, hybrid  resonances, exotic four-quark systems and pentaquarks
already attracted interests of physicists \cite{GellMann:1964nj,Jaffe:1977,Witten:1979kh,Balitsky:1982ps,Reinders:1985,
Braun:1985ah,Braun:1988kv,Jaffe:2004zg}. Besides general theoretical problems
of the multi-parton states, in some of these  works their  parameters were calculated, as well.
The $X$ resonances as the four-quark states can be treated within the diquark-antidiquark \cite{Jaffe:2004ph,Maiani:2004vq} or molecular pictures suggested to explain their
internal organization. In fact, in theoretical investigations of $X_1$ and $X_2$ both of these models were used: The resonances $X_1$ and $X_2$ were considered  as  the meson molecules in Refs.\ \cite{Liu:2008tn,Wang:2009ue,Albuquerque:2009ak,Wang:2009ry,Wang:2011uk,Liu:2010hf,He:2011ed, Finazzo:2011he,HidalgoDuque:2012pq} , whereas in Ref.\ \cite{Stancu:2009ka,Patel:2014vua} they were treated in the framework of the diquark-antidiquark model.
There are also alternative approaches analyzing them as dynamically generated
resonances \cite{Molina:2009ct,Branz:2010rj} or coupled-channel effects \cite{Danilkin:2009hr}. The recent comprehensive review of the various theoretical models, achieved progress and existing problems in the physics of multiquark resonances can be found in Ref.\ \cite{Esposito:2016noz}.

The experimental situation, stabilized after the LHCb report, imposes new
constraints on possible models of $X$ resonances. Indeed, an analysis
carried out by the LHCb Collaboration in Ref.\ \cite{Aaij:2016iza} on the
basis of the collected experimental information ruled out an explanation of
the $X_{1}$ as $0^{++}$ or $2^{++}$ $D_{s}^{\ast +}D_{s}^{\ast -}$ molecular
states. The LHCb also emphasized that molecular bound-states or cusps can
not account for $X_{2}$.

Therefore, in order to explain the experimental data, new models
and ideas are suggested. First of all, there are traditional attempts to
describe the $X$ resonances as excited states of the conventional charmonium
or as dynamical effects. Indeed, by analyzing experimental information of
the Belle and BaBar collaborations (see, Refs.\ \cite{Bhardwaj:2015rju} and
\cite{Aubert:2008bl}) on the $B\rightarrow K\chi _{c1}\pi ^{+}\pi ^{-}$ and $%
B\rightarrow KD\overline{D}$ decays, in Ref.\ \cite{Chen:2016iua} the author
identified the resonances $X_{1}$ and $Y(4080)$ with the P-wave excited
charmonium states $\chi _{c1}(3^{3}P_{1})$ and $\chi _{c0}(3^{3}P_{0})$,
respectively,

The contribution of the rescattering effects to the process $%
B^{+}\rightarrow J/\psi \phi K^{+}$ was studied in Ref.\ \cite{Liu:2016onn}
aiming to answer a question can they simulate the observed $X_{1}$, $X_{2}$,
$X_{3}$ and $X_{4}$ resonances or not. It was found that the $D_{s}^{\ast
+}D_{s}^{-}$ and $\psi ^{\prime }\phi $ rescatterings via meson loops may
simulate the structures $X_{1}$ and $X_{4}$, respectively. But, description
of the $X_{2}$ and $X_{3}$ states as rescattering effects seem are
problematic, which implies that they could be real four-quark resonances.
Nevertheless, on the basis of some other arguments (see, for details Ref.\
\cite{Liu:2016onn}) the author did not exclude treating of $X_{2}$ as the
excited $\chi _{c1}(3^{3}P_{1})$ state of the conventional charmonium.

The diquark-antidiquark and molecule-like models prevail other pictures and
form a theoretical basis for numerous calculations to account for available
information on the $X$ resonances \cite%
{Chen:2016oma,Chen:2010ze,Wang:2016tzr,Wang:2016dcb,Wang:2016gxp}. Thus, the
the masses of the axial-vector $J^{P}=1^{+}$ diquark-antidiquark $[cs][\bar c\bar s]$ states
with the triplet and sextet color structures were calculated in Ref.\  \cite{Chen:2010ze}.
Recently, in the light of the experimental data of the LHCb Collaboration, they were
interpreted as the $X_{1}$ and $X_{2}$ resonances, respectively \cite{Chen:2016oma}.
Within the same approach the $X_{3}$ and $X_{4}$ states were considered as the
$D$-wave excitations of the their light counterparts $X_1$ and $X_2$ \cite{Chen:2016oma}.

In the context of tetraquark models the resonances $X_1$ and $X_2$ were
studied in Refs.\ \cite{Wang:2016tzr} and \cite{Wang:2016dcb}, as well. In
accordance with Ref.\ \cite{Wang:2016tzr} the light $X_{1}$ resonance can
not be considered as the diquark-antidiquark compact state. The similar
conclusion was made in respect of $X_{2}$, which was examined as a
octet-octet type molecule-like state: The mass of the $X_2$ resonance found
there was in agreement with the LHCb data, but its decay width overshot
considerably the experimental result \cite{Wang:2016dcb}.  The scalar resonance
$X_{3}$ was considered as the first radial excitation of the axial-vector
 diquark-antidiquark $X(3915)$ state, whereas $X_{4} $ was analyzed as the ground
 state of the $[cs][\bar{c}\bar{s}]$ tetraquark built of the vector diquark and
 antidiquark  \cite{Wang:2016gxp}. Here
some comments about $X(3915)$ are in order. It was registered by the Belle
Collaboration as a resonance in the $J/\psi \omega $ invariant mass
distribution at the exclusive decay $B\rightarrow J/\psi \omega K$ \cite%
{Abe:2004zs}, and also seen in the reaction $\gamma \gamma \rightarrow
J/\psi \omega $ \cite{Uehara:2009tx}. This resonance was confirmed by the
BaBar Collaboration in the same $B\rightarrow J/\psi \omega K$ process \cite%
{Aubert:2007vj}. The $X(3915)$ was traditionally interpreted as the scalar $c%
\bar{c}$ meson $\chi _{c0}(2^{3}P_{0})$. But a lack of its expected $\chi
_{c0}(2P)\rightarrow D\overline{D} $ decay modes gave rise to other
conjectures. Thus, an alternative assumption concerning the $X(3915)$ resonance
was made in Ref.\ \cite{Lebed:2016yvr}, where it was identified with the
lightest scalar $[cs][\bar{c}\bar{s}]$ tetraquark state. Namely, this
resonance was considered in Ref.\ \cite{Wang:2016gxp} as the ground state of
$X_{3}$. Calculations seem confirm suggestions made on the nature of the $X_{3}$
and $X_{4}$ resonances \cite{Wang:2016gxp}.

An abundance of the observed charmonium-like resonances necessitated
spectroscopic analysis of the diquark-antidiquark states, which resulted in
suggestion of various multiplets to systemize  the discovered tetraquarks
(see, Refs.\ \cite{Stancu:2006st,Maiani:2016wlq,Zhu:2016arf}). The $X$ resonances were
 included into  $1S$ and $2S$ multiplets of color triplet $[cs]_{s=0,1}[\overline{c}%
\overline{s}]_{\overline{s}=0,1}$ tetraquarks \cite{Maiani:2016wlq}. Thus, $X_{1}$ was
identified with the $J^{PC}=1^{++}$ level of the $1S$ ground-state multiplet.
The $X_{2}$ resonance is supposedly, a linear superposition of two states with
$J^{PC}=0^{++}$ and $J^{PC}=2^{++}$. This suggestion was made, because in the multiplet
of the color triplet  tetraquarks only one state can bear the
quantum numbers $J^{PC}=1^{++}$. The heavy resonances $X_{3}$ and $X_{4}$ are
included into the $2S$ multiplet as its $J^{PC}=0^{++}$ members.
But apart from the color triplet multiplets there may  exist a multiplet
of the color sextet tetraquarks \cite{Stancu:2006st}, which also contains
a state with $J^{PC}=1^{++}$. In other words, the multiplet of the color sextet
tetraquarks doubles a number of the states with the same spin-parity \cite{Stancu:2006st},
and the $X_2$ resonance may be identified with its $J^{PC}=1^{++}$ member.

Even from this brief survey it is evident, that in the context of the
diquark-antidiquark model there exist different, sometimes contradictory
suggestions concerning the internal structure of the $X$ resonances.
Moreover, almost in all of these investigations the spectroscopic parameters
of newly discovered states were found by means of QCD two-point sum rule
method. Predictions of the sum rules for the parameters of the exotic states
extracted by using various assumptions on the interpolating currents, within
theoretical errors are consistent with the experimental data. In most of
cases results of various works are in accord with each other, as well. In
other words, the static parameters of the exotic states, such as their
masses, meson-current couplings are not enough to
verify existing models by confronting them with experimental data or/and
alternative theoretical models. The additional information useful in such
cases can be gained from investigation of decay channels of the exotic
states.

The QCD sum rule is the powerful nonperturbative method to explore the
exclusive hadronic processes and calculate parameters of hadrons, including
width of their strong decays \cite{Shifman:1979}. The width of the decay channels can be
computed by applying either the three-point sum rule approach or the
light-cone sum rule (LCSR) method \cite{Balitsky:1989ry}. The tetraquark
states dominantly decay to two conventional mesons. In the present work we
will study namely such decay modes of the $X_1$ and $X_2$ resonances.
Calculation of the couplings corresponding to strong vertices of a
tetraquark and two mesons in the context of the LCSR method requires usage
of additional technical tools. The reasons for a distinct treatment of
vertices with tetraquarks are very simple: Because these states are composed
of four valence quarks, the light-cone expansion of the relevant non-local
correlation function in terms of meson distribution amplitudes unavoidably
reduces to expressions with local matrix elements of the same meson. As a
result, conservation of the four-momentum in a such strong vertex is
fulfilled only if the four-momentum of this meson is set equal to zero. The
emerged situation can be handled by invoking into analysis technical tools,
known as a soft-meson approximation \cite{Ioffe:1983ju,Braun:1995}. For
investigation of the diquark-antidiquark states the soft-meson approximation
was adapted in Ref.\ \cite{Agaev:2016dev}, and successfully applied to
analyze decays some of the tetraquarks in Refs.\ \cite%
{Agaev:2016ijz,Agaev:2016lkl,Agaev:2016urs}.

In the present work we explore the properties of the $X_{1}$ and $X_{2}$
resonances in the context of QCD sum rule method. We are going to
interpolate $X_1$ and $X_2$, as in Ref.\ \cite{Chen:2010ze}, by the
spin-parity $J^{PC}=1^{++}$ currents with the antisymmetric and symmetric
color structures, respectively. By accepting this scheme we suggest that
there exist two different ground-state multiplets of triplet-triplet and
sextet-sextet type tetraquarks, and the $X_{1}$ and $X_{2}$ resonances are
their members with the same $J^{PC}=1^{++}$. Correctness of this hypothesis
can be checked by computing the masses of $X_{1}$ and $X_{2}$ states, and,
what is more important, their decay widths $\Gamma{(X_{1}\to J/\psi \phi})$
and $\Gamma({X_{2}\to J/\psi \phi})$. The masses and meson-current couplings
of $X_1$ and $X_2$ will computed by utilizing the two-point QCD sum rule
approach. We will also analyze the vertices $X_{1}J/\psi \phi $, $%
X_{2}J/\psi \phi $ and calculate the strong couplings $g_{X_{1}J/\psi \phi }$
and $g_{X_{2}J/\psi \phi }$ by means of the light-cone sum rule method
employing the soft-meson technique. Obtained results will enable us to find
the widths of the $X_{1}\rightarrow J/\psi \phi $ and $X_{2}\rightarrow
J/\psi \phi $ decays.

This work is structured in the following manner. In Sec.\ \ref{sec:Mass} we
calculate the masses and meson-current couplings of the $X_{1}$ and $X_{2}$
resonances.  In Sec.\ \ref{sec:Vertices} we find the
strong couplings corresponding to the vertices $X_{1}J/\psi \phi $ and $%
X_{2}J/\psi \phi $, and calculate the widths of the decay channels $%
X_{1}\rightarrow J/\psi \phi $ and $X_{2}\rightarrow J/\psi \phi $. In Sec.
\ref{sec:Conclusions} we compare our results with LHCb data and
predictions obtained in other works. It contains also our concluding remarks.
The explicit expressions of the quark propagators used in sum rule calculations
are moved to Appendix.


\section{Parameters of the $X(4140)$ and $X(4274)$ resonances}

\label{sec:Mass}
The QCD two-point sum rules for calculation of the masses and meson-current
couplings of the $X_{1}$ and $X_{2}$ resonances can be obtained from
analysis of the correlation function
\begin{equation}
\Pi _{\mu \nu }(q)=i\int d^{4}xe^{iq\cdot x}\langle 0|\mathcal{T}\{J_{\mu
}(x)J_{\nu }^{\dag }(0)\}|0\rangle ,  \label{eq:CorrF1}
\end{equation}%
where $J_{\mu }(x)$ is the interpolating current of the $X$ state with the
quantum numbers $J^{PC}=1^{++}$.

In accordance with the approach defended in Refs.\ \cite%
{Chen:2016oma,Chen:2010ze}, the $X_{1}$ and $X_{2}$ resonances have the same
quantum numbers, but different internal color organization. We follow their
assumptions and study the $X_{1}$ and $X_{2}$ states within QCD two-point
sum rule method using different interpolating currents. Namely, we suggest
that the current
\begin{eqnarray}
J_{\mu }^{1} &=&s_{a}^{T}C\gamma _{5}c_{b}\left( \overline{s}_{a}\gamma
_{\mu }C\overline{c}_{b}^{T}-\overline{s}_{b}\gamma _{\mu }C\overline{c}%
_{a}^{T}\right)  \notag \\
&&+s_{a}^{T}C\gamma _{\mu }c_{b}\left( \overline{s}_{a}\gamma _{5}C\overline{%
c}_{b}^{T}-\overline{s}_{b}\gamma _{5}C\overline{c}_{a}^{T}\right) ,
\label{eq:Curr1}
\end{eqnarray}%
which has the antisymmetric $\left[ \overline{3}_{c}\right] _{cs}\otimes %
\left[ 3_{c}\right] _{\overline{cs}}$ color structure, presumably describes
the resonance $X_{1}$, whereas
\begin{eqnarray}
J_{\mu }^{2} &=&s_{a}^{T}C\gamma _{5}c_{b}\left( \overline{s}_{a}\gamma
_{\mu }C\overline{c}_{b}^{T}+\overline{s}_{b}\gamma _{\mu }C\overline{c}%
_{a}^{T}\right)  \notag \\
&&+s_{a}^{T}C\gamma _{\mu }c_{b}\left( \overline{s}_{a}\gamma _{5}C\overline{%
c}_{b}^{T}+\overline{s}_{b}\gamma _{5}C\overline{c}_{a}^{T}\right) ,
\label{eq:Curr2}
\end{eqnarray}%
with the symmetric $\left[ 6_{c}\right] _{cs}\otimes \left[ \overline{6}_{c}%
\right] _{\overline{cs}}$ color organization corresponds to the tetraquark $%
X_{2}$. In Eqs.\ (\ref{eq:Curr1}) and (\ref{eq:Curr2}) $a$ and $b $ are
color indices, and $C$ is the charge conjugation matrix.

In order to derive required sum rules we find, as usual the expression of
the correlator in terms of the physical parameters of the $X$ state. To this
end, we saturate the correlation function with a complete set of states with
quantum numbers of $X$ and perform in Eq.\ (\ref{eq:CorrF1}) integration
over $x$ to get
\begin{equation}
\Pi _{\mu \nu }^{\mathrm{Phys}}(q)=\frac{\langle 0|J_{\mu }|X(q)\rangle
\langle X(q)|J_{\nu }^{\dagger }|0\rangle }{m_{X}^{2}-q^{2}}+...
\end{equation}%
with $m_{X}$ being the mass of the $X$ state. Here the dots indicate
contributions to the correlation function arising from the higher resonances
and continuum states. We introduce the meson-current coupling $f_{X}$ by
means of the matrix element
\begin{equation}
\langle 0|J_{\mu }|X(q)\rangle =f_{X}m_{X}\varepsilon _{\mu },
\label{eq:Res}
\end{equation}%
where $\varepsilon _{\mu }$ is the polarization vector of the $X$ resonance.
Then in terms of $m_{X}$ and $f_{X}$, the correlation function can be recast
to the form
\begin{equation}
\Pi _{\mu \nu }^{\mathrm{Phys}}(q)=\frac{m_{X}^{2}f_{X}^{2}}{m_{X}^{2}-q^{2}}%
\left( -g_{\mu \nu }+\frac{q_{\mu }q_{\nu }}{m_{X}^{2}}\right) +\ldots
\label{eq:CorM}
\end{equation}%
By applying the Borel transformation  to Eq.\ (\ref{eq:CorM}) we get
\begin{equation}
\mathcal{B}_{q^{2}}\Pi _{\mu \nu }^{\mathrm{Phys}%
}(q)=m_{X}^{2}f_{X}^{2}e^{-m_{X}^{2}/M^{2}}\left( -g_{\mu \nu }+\frac{q_{\mu
}q_{\nu }}{m_{X}^{2}}\right) +\ldots  \label{eq:CorBor}
\end{equation}

The QCD side of the sum rule has to be calculated by employing the
quark-gluon degrees of freedom. For this purpose, we contract the $c$ and $s$%
- quark fields and find for the correlation function $\Pi _{\mu \nu }^{%
\mathrm{QCD}}(q)$ the following expression (for definiteness, below we
provide explicit expression for the current $J_{\mu }^{1}$):
\begin{eqnarray}
&&\Pi _{\mu \nu }^{\mathrm{QCD}}(q)=-i\int d^{4}xe^{iqx}\epsilon \widetilde{%
\epsilon }\epsilon ^{\prime }\widetilde{\epsilon }^{\prime }\left\{ \mathrm{%
Tr}\left[ \gamma _{\mu }\widetilde{S}_{c}^{n^{\prime }n}(-x)\right. \right.
\notag \\
&&\left. \times \gamma _{\nu }S_{s}^{m^{\prime }m}(-x)\right] \mathrm{Tr}%
\left[ \gamma _{5}\widetilde{S}_{s}^{aa^{\prime }}(x)\gamma
_{5}S_{c}^{bb^{\prime }}(x)\right]  \notag \\
&&+\mathrm{Tr}\left[ \gamma _{\mu }\widetilde{S}_{c}^{n^{\prime
}n}(-x)\gamma _{5}S_{s}^{m^{\prime }m}(-x)\right] \mathrm{Tr}\left[ \gamma
_{\nu }\widetilde{S}_{s}^{aa^{\prime }}(x)\right.  \notag \\
&&\times \left. \gamma _{5}S_{c}^{bb^{\prime }}(x)\right] +\mathrm{Tr}\left[
\gamma _{5}\widetilde{S}_{c}^{n^{\prime }n}(-x)\gamma _{\nu
}S_{s}^{m^{\prime }m}(-x)\right]  \notag \\
&&\times \mathrm{Tr}\left[ \gamma _{5}\widetilde{S}_{s}^{aa^{\prime
}}(x)\gamma _{\mu }S_{c}^{bb^{\prime }}(x)\right] +\mathrm{Tr}\left[ \gamma
_{5}\widetilde{S}_{c}^{n^{\prime }n}(-x)\right.  \notag \\
&&\left. \times \left. \gamma _{5}S_{s}^{m^{\prime }m}(-x)\right] \mathrm{Tr}%
\left[ \gamma _{\nu }\widetilde{S}_{s}^{aa^{\prime }}(x)\gamma _{\mu
}S_{c}^{bb^{\prime }}(x)\right] \right\} ,  \label{eq:CorrF2}
\end{eqnarray}%
where $\epsilon =\epsilon ^{cab},\ \widetilde{\epsilon }=\epsilon ^{cmn}$
and $\epsilon ^{\prime }=\epsilon ^{c^{\prime }a^{\prime }b^{\prime }},\
\widetilde{\epsilon }^{\prime }=\epsilon ^{c^{\prime }m^{\prime }n^{\prime
}} $. In Eq.\ (\ref{eq:CorrF2}) $S_{s}^{ab}(x)$ and $S_{c}^{ab}(x)$ are the $%
s$ and $c$-quark propagators, respectively (see, Appendix \ref{sec:App}).
Here we also use the notation
\begin{equation}
\widetilde{S}_{s(c)}(x)=CS_{s(c)}^{T}(x)C.
\end{equation}

The QCD sum rule can be obtained by isolating the same Lorentz structures in
both of $\Pi _{\mu \nu }^{\mathrm{Phys}}(q)$ and $\Pi _{\mu \nu }^{\mathrm{%
QCD}}(q)$. We work with the terms $\sim g_{\mu \nu }$. The invariant
amplitude $\Pi ^{\mathrm{QCD}}(q^{2})$ corresponding to this structure can
be written down as the dispersion integral
\begin{equation}
\Pi ^{\mathrm{QCD}}(q^{2})=\int_{4(m_{c}+m_{s})^{2}}^{\infty }\frac{\rho ^{%
\mathrm{QCD}}(s)}{s-q^{2}}ds+...,
\end{equation}%
where $\rho ^{\mathrm{QCD}}(s)$ is the two-point spectral density. By
applying the Borel transformation to $\Pi ^{\mathrm{QCD}}(q^{2})$ , equating
the obtained expression with the relevant part of the function $\mathcal{B}%
_{q^{2}}\Pi _{\mu \nu }^{\mathrm{Phys}}(q)$, and subtracting the continuum
contribution we find the final sum rule. The mass of the $X$ state can be
evaluated from the sum rule
\begin{equation}
m_{X}^{2}=\frac{\int_{4(m_{c}+m_{s})^{2}}^{s_{0}}dss\rho ^{\mathrm{QCD}%
}(s)e^{-s/M^{2}}}{\int_{4(m_{c}+m_{s})^{2}}^{s_{0}}ds\rho (s)e^{-s/M^{2}}},
\label{eq:srmass}
\end{equation}%
whereas to find the meson-current coupling $f_{X}$ we employ the expression
\begin{equation}
f_{X}^{2}m_{X}^{2}e^{-m_{X}^{2}/M^{2}}=\int_{4(m_{c}+m_{s})^{2}}^{s_{0}}ds%
\rho ^{\mathrm{QCD}}(s)e^{-s/M^{2}}.  \label{eq:srcoupling}
\end{equation}
\begin{table}[tbp]
\begin{tabular}{|c|c|}
\hline\hline
Parameters & Values \\ \hline\hline
$m_{J/\psi}$ & $(3096.900 \pm 0.006) ~\mathrm{MeV}$ \\
$f_{J/\psi}$ & $405~\mathrm{MeV}$ \\
$m_{\phi}$ & $(1019.461 \pm 0.019) ~\mathrm{MeV}$ \\
$f_{\phi}$ & $215 \pm 5 ~\mathrm{MeV}$ \\
$m_{c}$ & $(1.27 \pm 0.03)~\mathrm{GeV}$ \\
$m_{s} $ & $96^{+8}_{-4}~\mathrm{MeV} $ \\
$\langle \bar{q}q \rangle $ & $-(0.24\pm 0.01)^3$ $\mathrm{GeV}^3$ \\
$\langle \bar{s}s \rangle $ & $0.8\ \langle \bar{q}q \rangle$ \\
$m_{0}^2 $ & $(0.8\pm0.1)$ $\mathrm{GeV}^2$ \\
$\langle \overline{s}g_{s}\sigma Gs\rangle$ & $m_{0}^2\langle \bar{s}s
\rangle $ \\
$\langle\frac{\alpha_sG^2}{\pi}\rangle $ & $(0.012\pm0.004)$ $~\mathrm{GeV}%
^4 $ \\
$\langle g_{s}^3G^3\rangle $ & $(0.57\pm0.29)$ $~\mathrm{GeV}^6 $ \\
\hline\hline
\end{tabular}%
\caption{Parameters used in sum rule calculations.}
\label{tab:Param}
\end{table}


\begin{widetext}

\begin{figure}[h!]
\begin{center}
\includegraphics[totalheight=6cm,width=8cm]{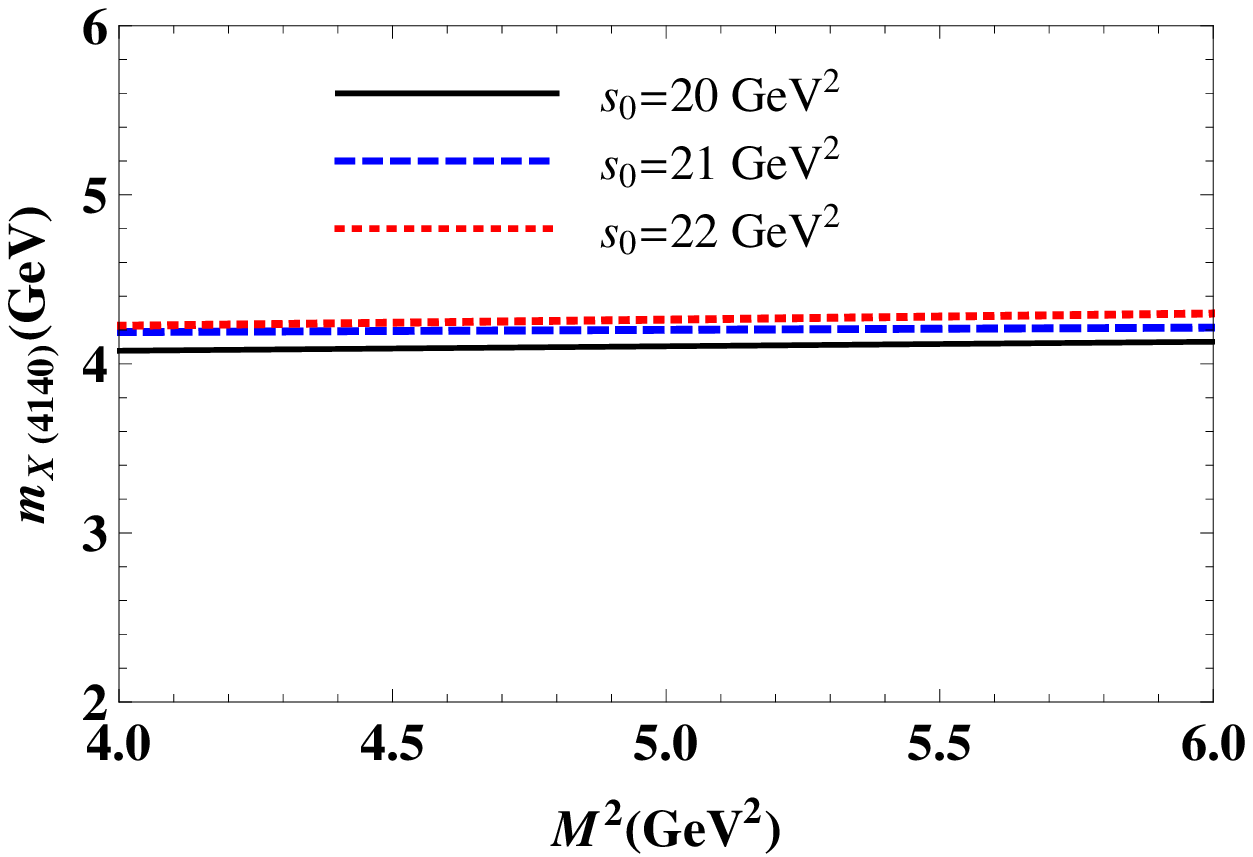}\,\, %
\includegraphics[totalheight=6cm,width=8cm]{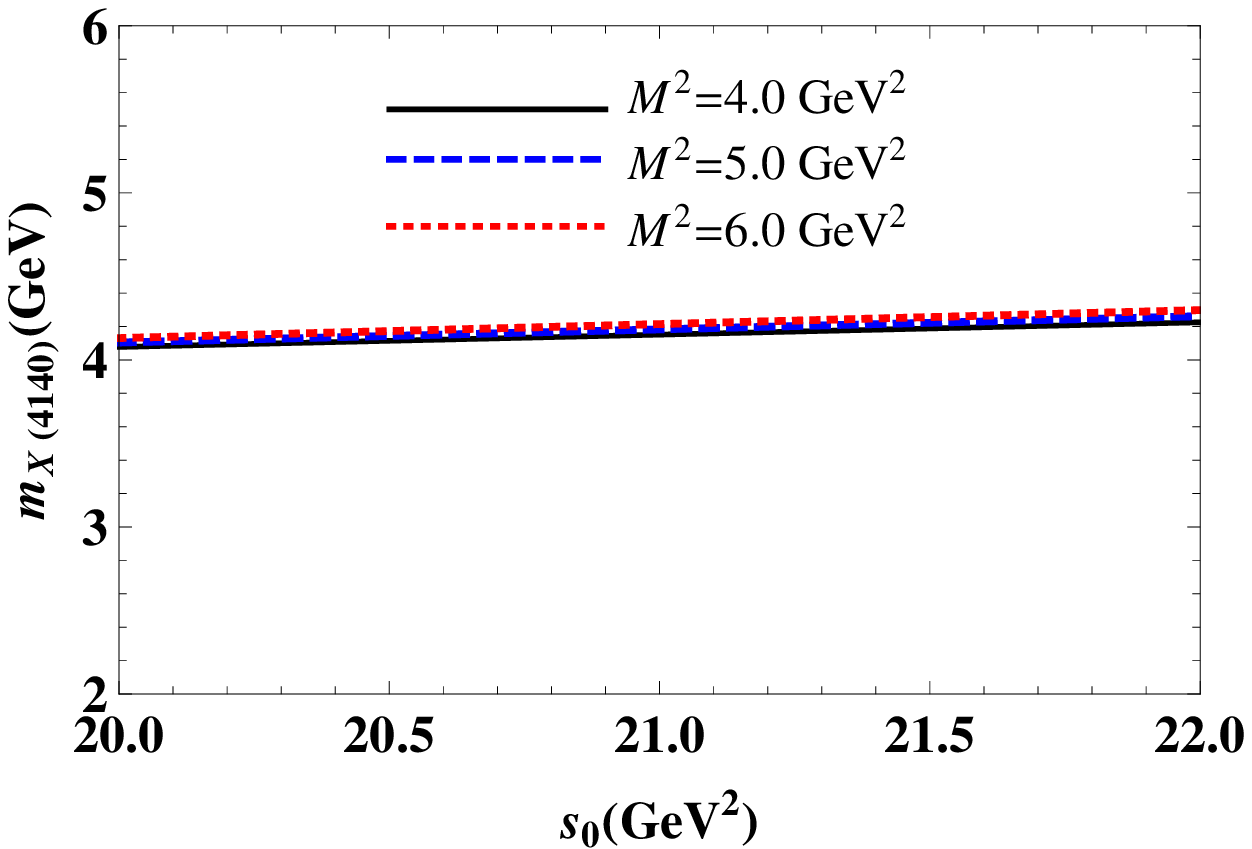}
\end{center}
\caption{ The mass of the $X(4140)$ state as a function of the Borel parameter $M^2$ at fixed $s_0$ (left panel), and as a function of the continuum threshold $s_0$ at fixed $M^2$ (right panel).}
\label{fig:Mass}
\end{figure}
\begin{figure}[h!]
\begin{center}
\includegraphics[totalheight=6cm,width=8cm]{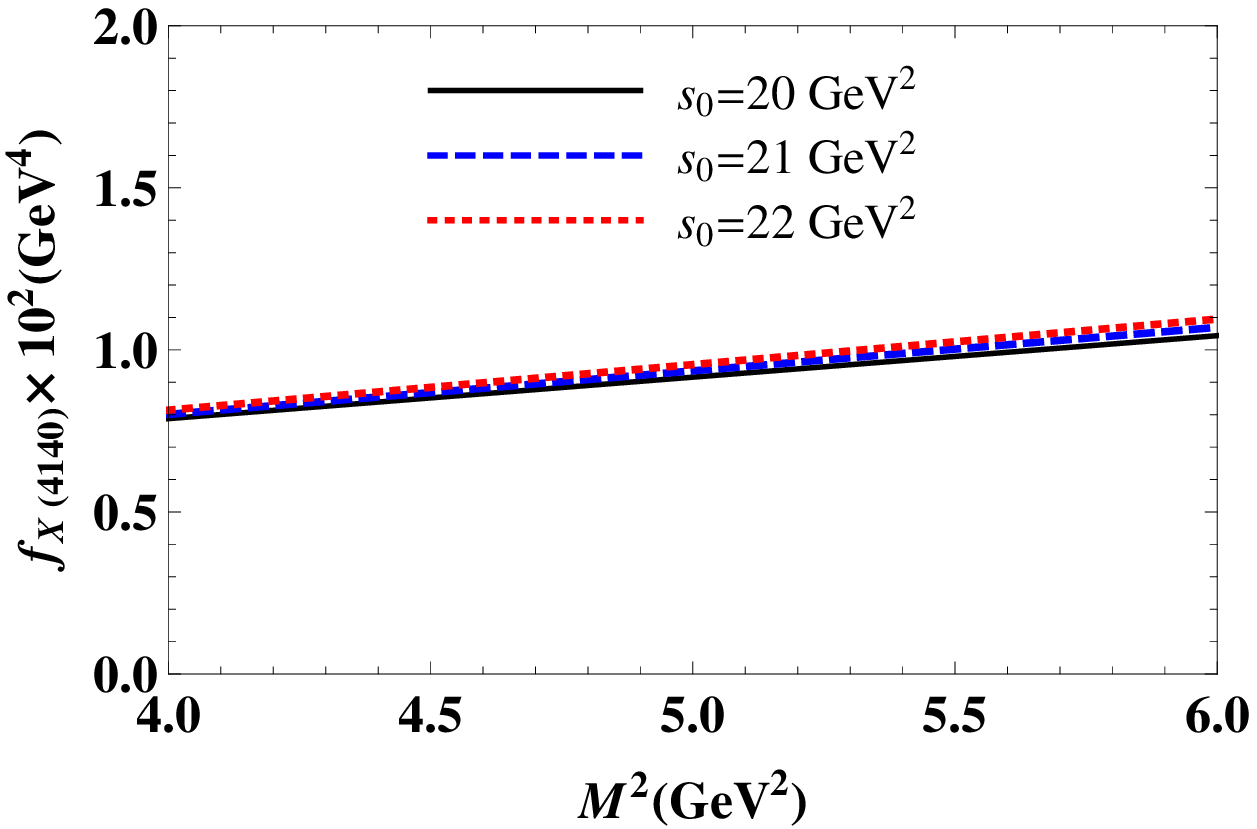}\,\, %
\includegraphics[totalheight=6cm,width=8cm]{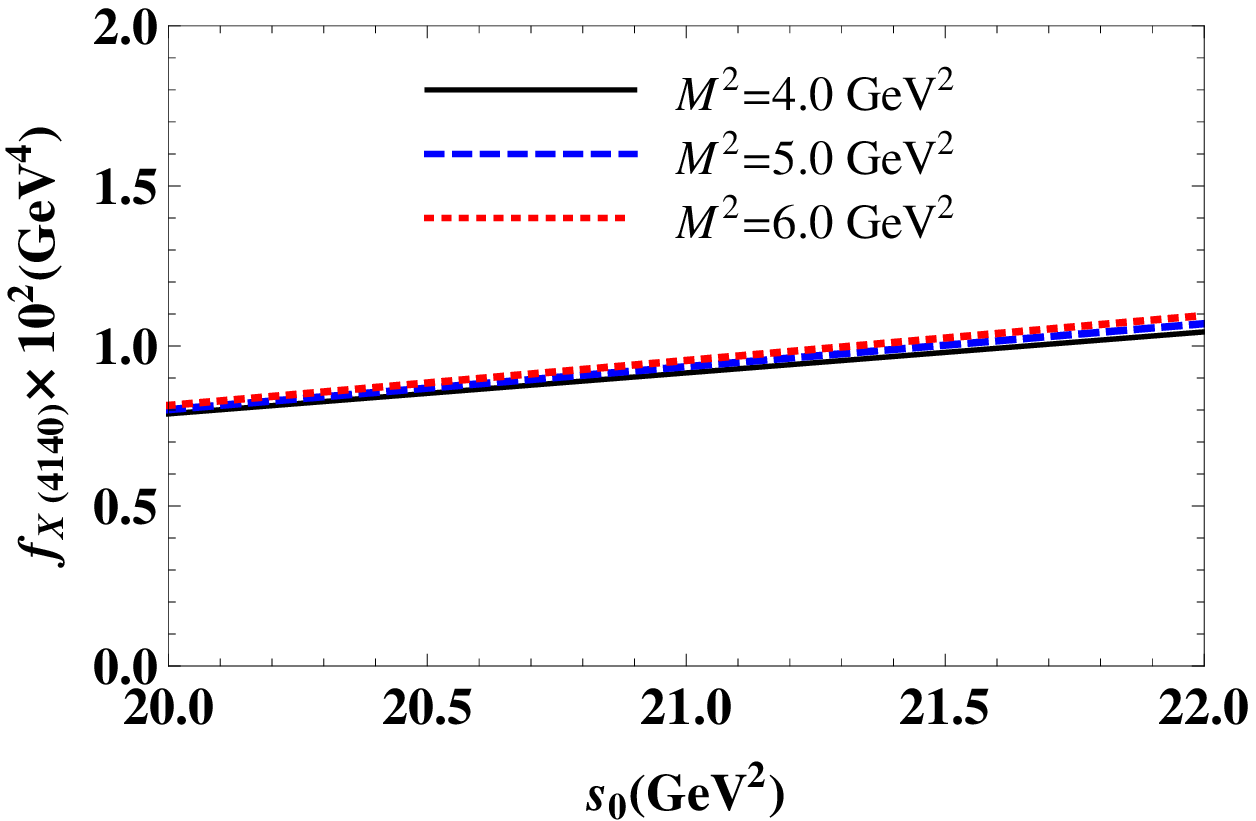}
\end{center}
\caption{ The dependence of the meson-current coupling $f_X$ of the $X(4140)$ resonance on the Borel parameter at chosen values of $s_0$ (left panel), and on the $s_0$ at fixed $M^2$  (right panel).}
\label{fig:Coup}
\end{figure}
\begin{figure}[h!]
\begin{center}
\includegraphics[totalheight=6cm,width=8cm]{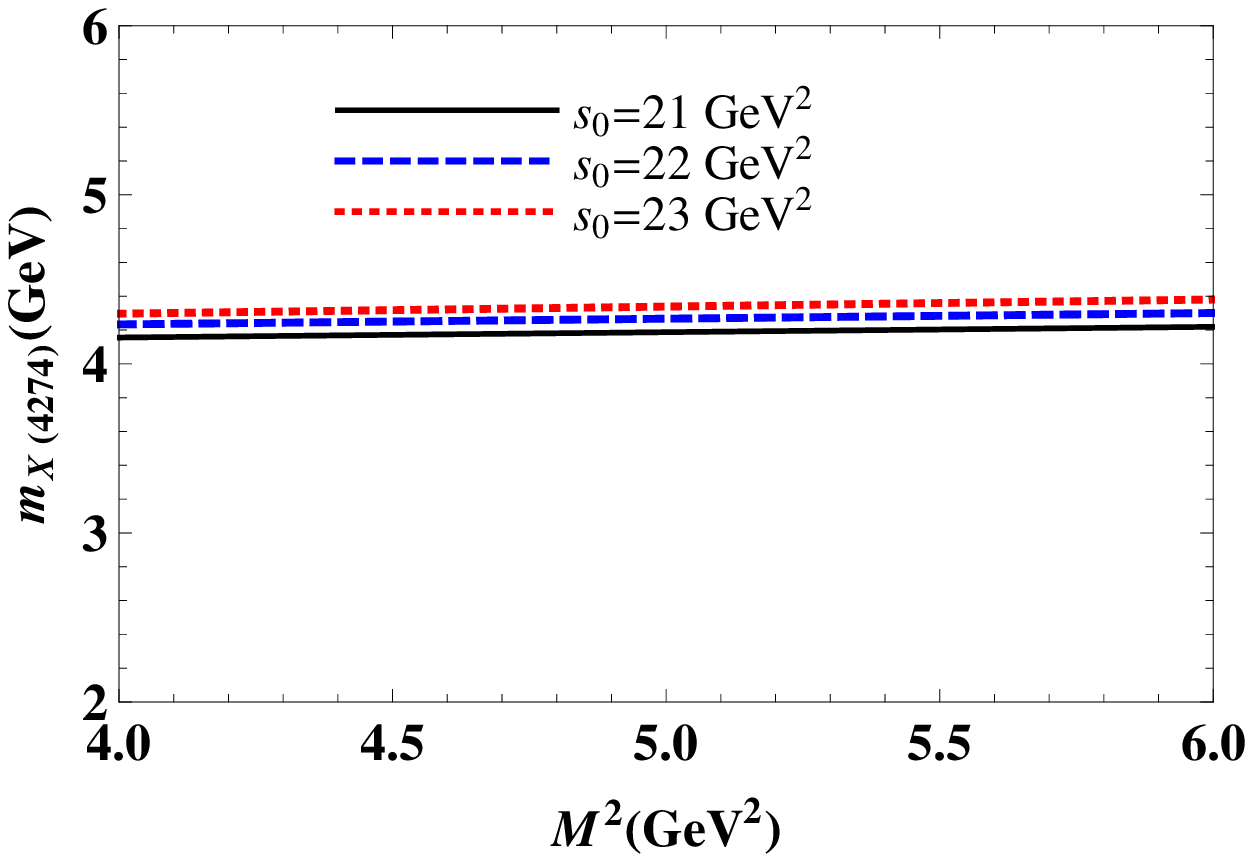}\,\, %
\includegraphics[totalheight=6cm,width=8cm]{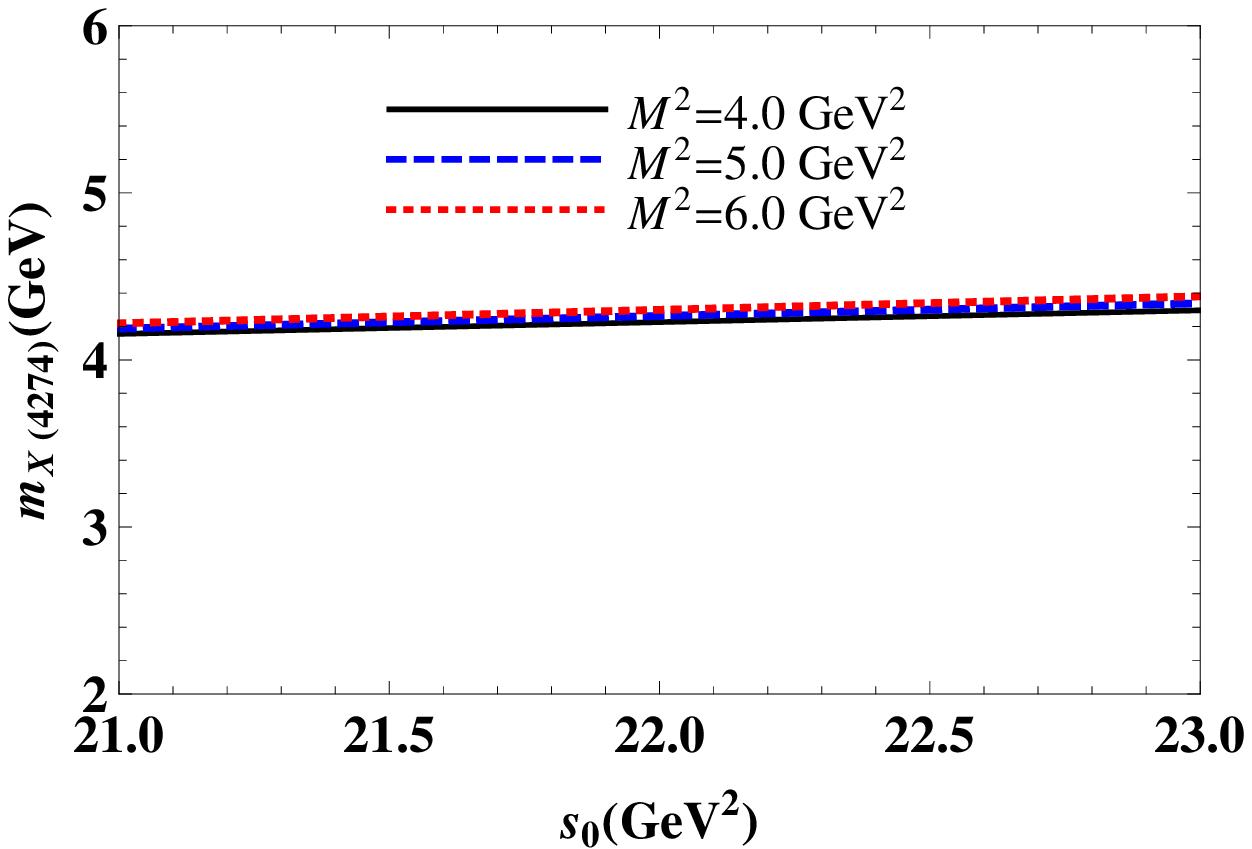}
\end{center}
\caption{ The mass of the $X(4274)$ resonance as a function of the Borel parameter $M^2$ at fixed $s_0$ (left panel), and as a function of the continuum threshold $s_0$ at fixed $M^2$ (right panel).}
\label{fig:Mass2}
\end{figure}
\begin{figure}[h!]
\begin{center}
\includegraphics[totalheight=6cm,width=8cm]{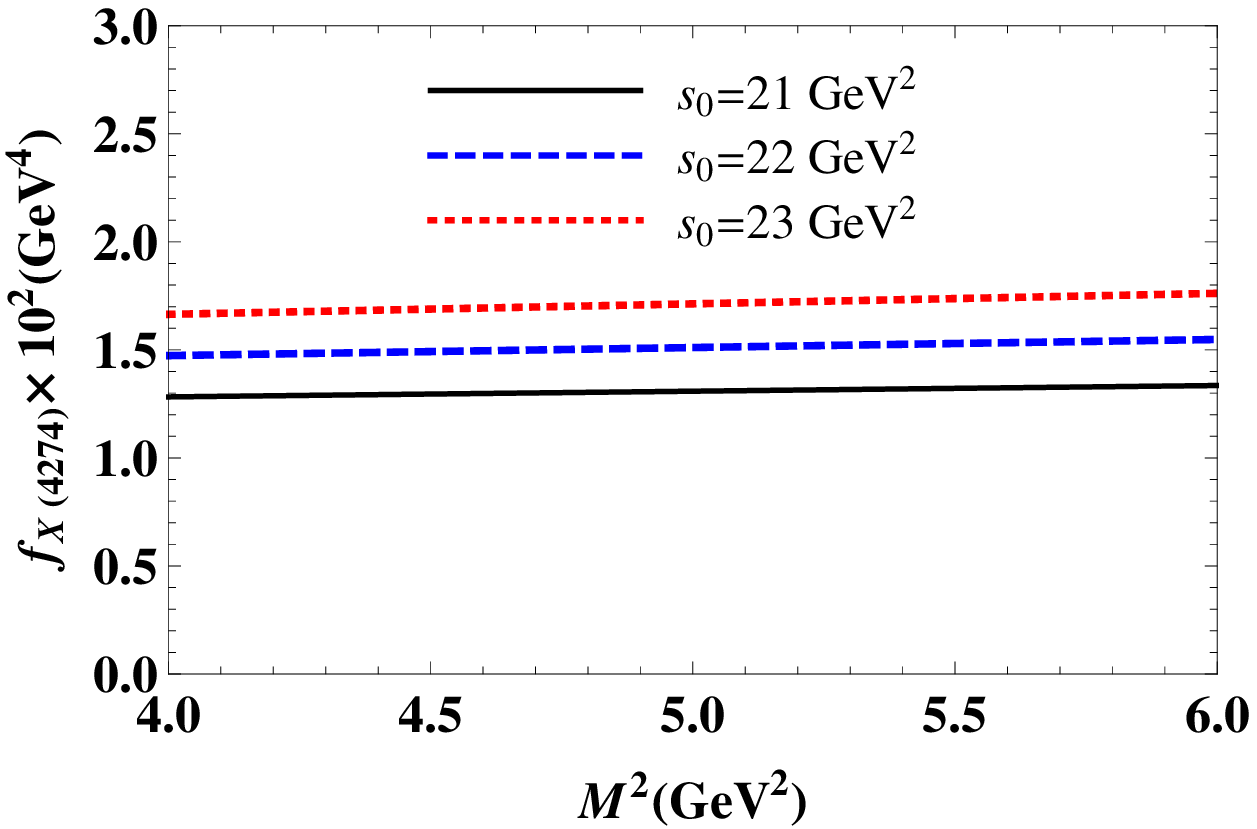}\,\, %
\includegraphics[totalheight=6cm,width=8cm]{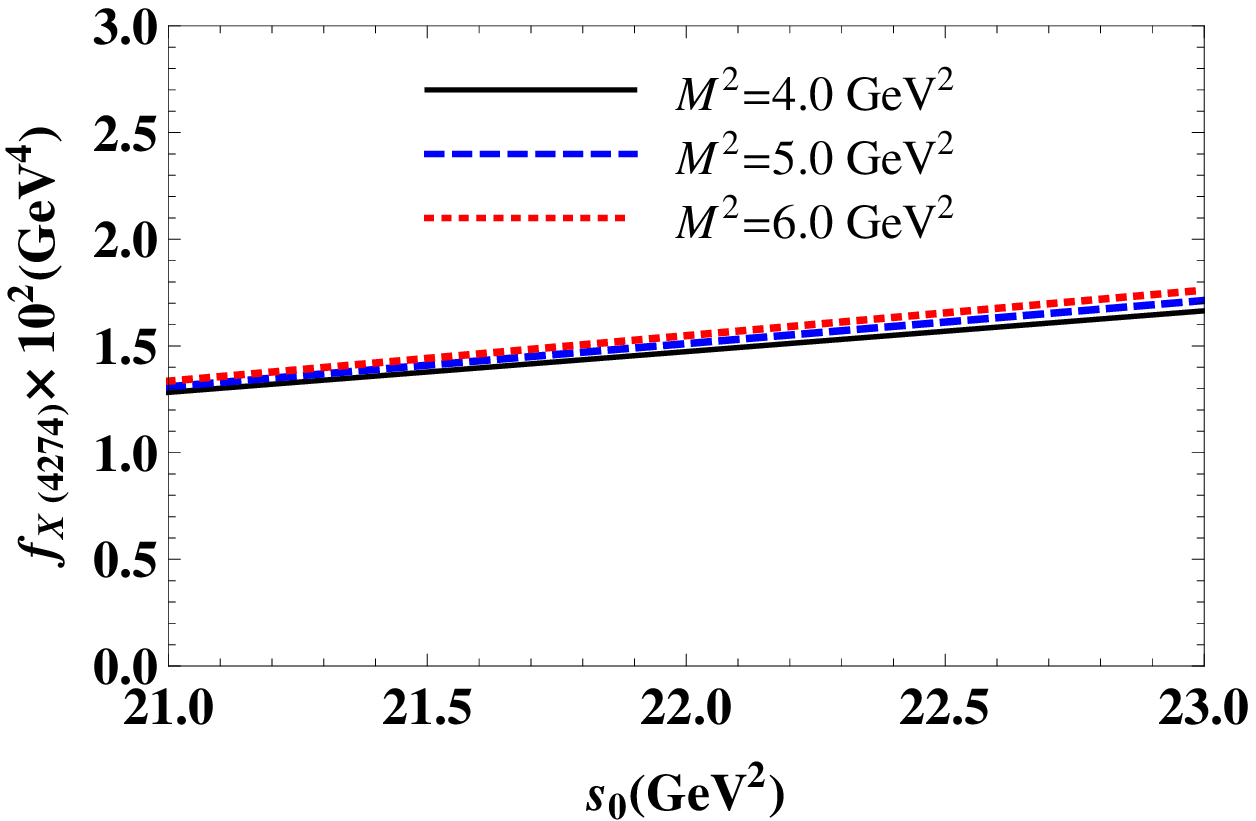}
\end{center}
\caption{ The meson-current coupling $f_X$ of the $X(4140)$ resonance as a function of the Borel parameter $M^2$ at chosen values of $s_0$ (left panel), and as a function of $s_0$ at fixed $M^2$  (right panel).}
\label{fig:Coup2}
\end{figure}

\end{widetext}

\begin{table}[tbp]
\begin{tabular}{|c|c|c|}
\hline\hline
$X$ & $X(4140)$ & $X(4274)$ \\ \hline\hline
$M^2 ~(\mathrm{GeV}^2$) & $4-6$ & $4-6$ \\ \hline
$s_0 ~(\mathrm{GeV}^2$) & $20-22$ & $21-23$ \\ \hline
$m_{X} ~(\mathrm{MeV})$ & $4183\pm 115$ & $4264 \pm 117$ \\ \hline
$f_{X} ~(\mathrm{GeV}^4)$ & $(0.94 \pm 0.16)\cdot 10^{-2}$ & $(1.51 \pm
0.21)\cdot 10^{-2}$ \\ \hline\hline
\end{tabular}%
\caption{The masses and meson-current couplings of the $X(4140)$ and $%
X(4274) $ tetraquarks.}
\label{tab:Results1A}
\end{table}
The methods for deriving of the spectral density $\rho ^{\mathrm{QCD}}(s)$
were presented in the literature (see, for example, Ref.\ \cite%
{Agaev:2016dev}. Therefore, we do not concentrate here on details of these
standard and rather routine calculations.

The expressions for the mass and meson-current coupling given by Eqs.\ (\ref%
{eq:srmass}) and (\ref{eq:srcoupling}) contain the input parameters,
numerical values of which are collected in Table \ref{tab:Param}. The sum
rules depend also on the auxiliary parameters $M^2$ and $s_0$. In general,
physical quantities extracted from the sum rules should not depend on the
Borel parameter and continuum threshold, but in real calculations we can
only minimize their effect on obtained results. They have also to obey the
standard requirements imposed by the sum rule calculations. Thus, in the
working regions of these parameters a prevalence of the pole contribution to
the sum rules and convergence of the operator product expansion (OPE) have
to be satisfied. Namely these restrictions, and a stability of the obtained
predictions determine ranges within of which the parameters $M^2$ and $s_0$
can be varied. Results of our analysis are collected in Table \ref%
{tab:Results1A}, where we provide both the working windows for the
parameters $M^2$ and $s_0$, as well as, the sum rule's results for the mass
and meson-current couplings of the $X(4140)$ and $X(4274)$ resonances. In
the working ranges of the parameters the pole contributions equal to $ 23\%$
of the whole results, which are typical for the sum rule calculations
involving four-quark systems. In order to control the convergence of OPE we
evaluate the contribution arising from each term of the fixed dimension: in
the ranges shown in Table \ref{tab:Results1A} convergence of OPE is
fulfilled: It is enough to note that contribution of the dimension-$8$ term
to the final result does not exceed $ 1\%$ of its value.

As is seen from Figs.\ \ref{fig:Mass} and \ref{fig:Coup}, the mass and
meson-current coupling of the $X(4140)$ state are sensitive to the
parameters $M^2$ and $s_0$: While their effects on the mass $m_X$ are mild,
the dependence of the meson-current coupling $f_X$ on the chosen values of
the Borel and continuum threshold parameters is noticeable. These effects
combined with ambiguities of the input parameters generate the theoretical
errors in the sum rule calculations, which are their unavoidable feature.
The errors of the calculations are also presented in Table \ref%
{tab:Results1A}. The similar estimations are valid for the $X(4274)$ state,
as well (see Figs.\ \ref{fig:Mass2} and \ref{fig:Coup2}).

The masses of the $X(4140)$ and $X(4274)$ found in the present work, are in
a nice agreement with LHCb data. At this stage of our investigations we can
conclude that $X(4140)$ and $X(4274)$  are the diquark-antidiquark $J^{PC}=1^{++}$
states of the color triplet and sextet multiplets, respectively.


\section{ Width of $X(4140)\rightarrow J/\protect\psi \protect\phi $ \ and $%
X(4274)\rightarrow J/\protect\psi \protect\phi $ decays}

\label{sec:Vertices}
The $X_{1}$ and $X_{2}$ $\ $states were observed as resonances in the $%
J/\psi \phi $ invariant mass distribution. Therefore, processes $%
X_{1}\rightarrow J/\psi \phi $ and $X_{2}\rightarrow J/\psi \phi $ may be
considered as their main decays channels. In this section we are going to
concentrate namely on these two decay processes. We will outline steps
necessary to analyze the vertex $XJ/\psi \phi $, where $X$ is one of the $%
X_1 $ and $X_2$ states, and calculate the strong coupling $g_{XJ/\psi \phi }$
and width of the decay $X\to J/\psi\phi$.

Within the sum rule method the strong vertex\ $XJ/\psi \phi $ can be studied
using the correlation function
\begin{equation}
\Pi _{\mu \nu }(p,q)=i\int d^{4}xe^{ipx}\langle \phi (q)|\mathcal{T}\{J_{\mu
}^{J/\psi }(x)J_{\nu }^{\dag }(0)\}|0\rangle ,  \label{eq:CorrF3}
\end{equation}%
where $J_{\nu }$ and $J_{\mu }^{J/\psi }$ are the interpolating currents of
the $X$ state and $J/\psi $ meson, respectively. The current $J_{\nu }$ is
defined by one of Eqs.\ (\ref{eq:Curr1}) and (\ref{eq:Curr2}), whereas $%
J/\psi $ has the form
\begin{equation}
J_{\mu }^{J/\psi }(x)=\overline{c}_{l}(x)\gamma _{\mu }c_{l}(x).
\label{eq:Bcur}
\end{equation}%
We calculate $\Pi _{\mu \nu }(p,q)$ employing QCD sum rule on the light-cone
and soft approximation. To this end, at first stage of calculations one has
to express this function in terms of the physical quantities, namely in
terms of the masses, decay constants of involved particles, and strong
coupling $g_{XJ/\psi \phi }$ itself. For $\Pi _{\mu\nu }^{\mathrm{Phys}%
}(p,q) $ we get
\begin{eqnarray}
\Pi _{\mu\nu }^{\mathrm{Phys}}(p,q) &=&\frac{\langle 0|J_{\mu }^{J/\psi
}|J/\psi \left( p\right) \rangle }{p^{2}-m_{J/\psi }^{2}}\langle J/\psi
\left( p\right) \phi (q)|X(p^{\prime })\rangle  \notag \\
&&\times \frac{\langle X(p^{\prime })|J_{\nu }^{\dagger }|0\rangle }{%
p^{\prime 2}-m_{X}^{2}}+\ldots ,  \label{eq:CorrF4}
\end{eqnarray}%
where $p$, $q$ are the momenta of the $J/\psi $ and $\phi $ mesons,
respectively, and by $p^{\prime }=p+q$ we denote the momentum of the $X$
state.

We define the matrix element of the $J/\psi $ meson in the form
\begin{equation*}
\langle 0|J_{\mu }^{J/\psi }|J/\psi \left( p\right) \rangle =f_{J/\psi
}m_{J/\psi }\varepsilon _{\mu }(p),
\end{equation*}%
where $m_{J/\psi }$, $f_{J/\psi }$ and $\varepsilon _{\mu }(p)$ are its
mass, decay constant and polarization vector, respectively. We introduce
also the matrix element corresponding to the vertex
\begin{eqnarray}
&&\langle J/\psi \left( p\right) \phi (q)|X(p^{\prime })\rangle   \notag \\
&&=ig_{XJ/\psi \phi }\epsilon_{\alpha \beta \gamma \delta }\varepsilon
_{\alpha }^{\ast }(p)\varepsilon _{\beta }(p^{\prime })\varepsilon _{\gamma
}^{\ast }(q)p_{\delta }.  \label{eq:Mel}
\end{eqnarray}%
Here $\varepsilon _{\gamma }^{\ast }(q)$ is the polarization vector of the $%
\phi $ meson. Then the contribution coming from the ground state takes the
form
\begin{eqnarray}
&&\Pi _{\mu \nu }^{\mathrm{Phys}}(p,q)=i\frac{f_{J/\psi }f_{X}m_{J/\psi
}m_{X}g_{XJ/\psi \phi }}{\left( p^{\prime 2}-m_{X}^{2}\right) \left(
p^{2}-m_{J/\psi }^{2}\right) }  \notag \\
&&\times \left( \epsilon_{\mu \nu \gamma \delta }\varepsilon _{\gamma
}^{\ast }(p)p_{\delta }-\frac{1}{m_{X}^{2}}\epsilon _{\mu \beta \gamma
\delta }\varepsilon _{\gamma }^{\ast }(p)p_{\delta }p_{\beta }^{\prime
}p_{\nu }^{\prime }\right) +\ldots \notag \\
&&{} \label{eq:CorrF5}
\end{eqnarray}%
In the soft limit $p=p^{\prime }$ (see, a discussion below
and  Ref.\ \cite{Agaev:2016dev}), and only the term
$\sim i\epsilon _{\mu \nu \gamma \delta}\varepsilon _{\gamma }^{\ast }(p)p_{\delta }$
survives in Eq.\ (\ref{eq:CorrF5}).

In the soft-meson approximation we employ the one-variable Borel
transformation on $p^{2}$. Then, an invariant amplitude $\Pi ^{\mathrm{Phys}%
}(p^{2})$ depends on the variable $p^{2}$
\begin{equation}
\Pi ^{\mathrm{Phys}}(p^{2})=\frac{f_{J/\psi }f_{X}m_{J/\psi }m_{X}g_{XJ/\psi
\phi }}{\left( p^{2}-m^{2}\right) ^{2}},
\end{equation}%
where $m^{2}=(m_{X}^{2}+m_{J/\psi }^{2})/2$. Additionally, we apply to both
sides of the sum rule the operator
\begin{equation}
\left( 1-M^{2}\frac{d}{dM^{2}}\right) M^{2}e^{m^{2}/M^{2}},
\label{eq:softop}
\end{equation}%
which eliminates effects of unsuppressed terms in $\Pi ^{\mathrm{Phys}%
}(p^{2})$ appeared in the soft limit \cite{Ioffe:1983ju,Braun:1995}.

The QCD expression for the correlation function $\Pi _{\mu \nu }^{\mathrm{QCD%
}}(p,q)$ is calculated employing the quark propagators. For the current $%
J_{\mu}^{1}$ it takes the following form
\begin{eqnarray}
&&\Pi _{\mu \nu }^{\mathrm{QCD}}(p,q)=i\int d^{4}xe^{ipx}\epsilon
^{ijk}\epsilon ^{imn}  \notag \\
&&\times \left\{ \left[ \gamma _{\nu }\widetilde{S}_{c}^{ak}(x)\gamma _{\mu }%
\widetilde{S}_{c}^{na}(-x){}\gamma _{5}\right] \right.  \notag \\
&&\left. -\left[ \gamma _{5}\widetilde{S}_{c}^{ak}(x){}\gamma _{\mu }%
\widetilde{S}_{c}^{na}(-x){}\gamma _{\nu }\right] \right\} _{\alpha \beta
}\langle \phi (q)|\overline{s}_{\alpha }^{j}s_{\beta }^{m}|0\rangle ,
\label{TranCF1}
\end{eqnarray}%
with $\alpha $ and $\beta $ being the spinor indices.

To proceed we employ the replacement
\begin{equation}
\overline{s}_{\alpha }^{j}s_{\beta }^{m}\rightarrow \frac{1}{4}\Gamma
_{\beta \alpha }^{k}\left( \overline{s}^{j}\Gamma ^{k}s^{m}\right) ,
\label{eq:MatEx}
\end{equation}%
where $\Gamma ^{k}$ is the full set of Dirac matrices, and carry out the
color summation. Then we substitute Eq.\ (\ref{eq:Qprop}) into the
expression obtained after the color summation and perform four dimensional
integration over $x$. This integration leads to appearance of the Dirac
delta $\delta^4(p^{\prime}-p)$ in the integrand. The correlation function
does not contain the $s$-quark propagator, therefore the argument of the
Dirac delta depends only on the four-momenta of the $X$ state and $J/\psi$
meson. The next operation, i.e. an integration over $p$ or $p^{\prime}$
inevitably equates $p=p^{\prime}$, which is the result of the conservation
of the four-momentum at the vertex $XJ/\psi\phi$. In other words, to
conserve the four-momentum in the tetraquark-meson-meson vertex one should
set $q=0$, which in the full LCSR is known as the soft-meson
approximation \cite{Braun:1995}. At vertices of conventional mesons,
in general $q\neq 0$, and only in the soft-meson approximation one sets
$q$ equal to zero, whereas the tetraquark-meson-meson vertex can be treated
in the context of the LCSR method only if $q=0$. Nevertheless, an important
observation made in Ref.\ \cite{Braun:1995} is that,
both the soft-meson approximation and full LCSR treatment of the  ordinary mesons'
vertices lead for the strong couplings to very close numerical results.

In the soft limit only the matrix element
\begin{equation}
\langle 0|\overline{s}(0)\gamma _{\mu }s(0)|\phi (p,\lambda )\rangle
=f_{\phi }m_{\phi }\epsilon _{\mu }^{(\lambda )},
\end{equation}%
of the $\phi $- meson contributes to the correlation function, where $%
m_{\phi }$ and $f_{\phi }$ are its mass and decay constant, respectively.
The soft-meson limit reduces also possible Lorentz structures in $\Pi _{\mu
\nu }^{\mathrm{QCD}}(p,q)$ to the term $\sim i\epsilon _{\mu \nu \gamma
\delta }\varepsilon _{\gamma }^{\ast }(p)p_{\delta }$, which matches with
the corresponding structure in $\Pi _{\mu \nu }^{\mathrm{Phys}}(p,q=0)$.

\begin{table}[tbp]
\begin{tabular}{|c|c|c|}
\hline\hline
$X$ & $X(4140)$ & $X(4274)$ \\ \hline\hline
$M^2 ~(\mathrm{GeV}^2$) & $5-7$ & $5-7$ \\ \hline
$s_0 ~(\mathrm{GeV}^2$) & $20-22$ & $21-23$ \\ \hline
$g_{XJ/\psi \phi}$ & $2.34 \pm 0.89$ & $3.41 \pm 1.21$ \\ \hline
$\Gamma (X \to J/\psi \phi) ~(\mathrm{MeV})$ & $80 \pm 29$ & $272 \pm 81$ \\
\hline\hline
\end{tabular}%
\caption{The strong coupling $g_{XJ/\protect\psi \protect\phi}$ and decay
width $\Gamma (X \to J/\protect\psi \protect\phi)$.}
\label{tab:Results2A}
\end{table}

The relevant invariant amplitude can be written down as a dispersion
integral in terms of the spectral density $\rho _{c}^{\mathrm{QCD}}(s)$. We
omit details of calculations and provide the final expression for $\rho
_{c}^{\mathrm{QCD}}(s)$, which read
\begin{equation}
\rho _{c}^{\mathrm{QCD}}(s)=\frac{f_{\phi }m_{\phi }m_{c}}{4}\left[ \frac{%
\sqrt{s(s-4m_{c}^{2})}}{\pi ^{2}s}+\digamma ^{\mathrm{n.-pert.}}(s)\right].
\label{eq:Sdensity}
\end{equation}%
The nonperturbative component of $\rho_{c}^{\mathrm{QCD}}(s)$, i.e. $%
\digamma ^{\mathrm{n.-pert.}}(s)$ is given by the following formula
\begin{eqnarray}
&&\digamma ^{\mathrm{n.-pert.}}(s)=\Big \langle\frac{\alpha _{s}G^{2}}{\pi }%
\Big \rangle\int_{0}^{1}f_{1}(z,s)dz+\Big \langle g_{s}^{3}G^{3}\Big \rangle
\notag \\
&&\times \int_{0}^{1}f_{2}(z,s)dz+\Big \langle\frac{\alpha _{s}G^{2}}{\pi }%
\Big \rangle^{2}\int_{0}^{1}f_{3}(z,s)dz,  \label{eq:NPert}
\end{eqnarray}%
where the terms proportional to $\langle \alpha _{s}G^{2}/\pi \rangle $, $%
\langle g_{s}^{3}G^{3}\rangle $ and $\langle \alpha _{s}G^{2}/\pi \rangle
^{2}$ are nonperturbative contributions to the spectral density and have
four, six and eight dimensions, respectively. The explicit form of the
functions $f_{1}(z,s),\ f_{2}(z,s)$ and $f_{3}(z,s)$ are:
\begin{eqnarray}
&&f_{1}(z,s)=\frac{1}{18r^{2}}\left\{ -\left( 2+3r(3+2r)\right) \delta
^{(1)}(s-\Phi )\right.  \notag \\
&&\left. +(1+2r)\left[ m_{c}^{2}-sr\right] \delta ^{(2)}(s-\Phi )\right\}
,
\end{eqnarray}%
\begin{eqnarray}
&&f_{2}(z,s)=\frac{(1-2z)}{2^{7}\cdot 9\pi ^{2}r^{5}}\left\{ 2r\left[
3r\left( 1+rR\right) \delta ^{(2)}(s-\Phi )\right. \right.  \notag \\
&&\left. +\left[ 3sr^{2}(1+r)-2m_{c}^{2}\left( 1+rR\right) \right] \delta
^{(3)}(s-\Phi )\right] +  \notag \\
&&+\left[ s^{2}r^{4}-2sm_{c}^{2}r^{2}(1+r)+m_{c}^{4}(1+rR)\right]  \notag \\
&&\left. \times \delta ^{(4)}(s-\Phi )\right\} ,
\end{eqnarray}%
\begin{equation}
f_{3}(z,s)=\frac{m_{c}^{2}\pi ^{2}}{2^{2}\cdot 3^{4}r^{2}}\left[ \delta
^{(4)}(s-\Phi )-s\delta ^{(5)}(s-\Phi )\right] ,
\end{equation}%
where we introduce the short-hand notations%
\begin{equation}
\ r=z(z-1),\ R=3+r,\ \ \Phi =\frac{m_{c}^{2}}{z(1-z)},
\end{equation}%
and $\ \delta ^{(n)}(s-\Phi )$ is defined as
\begin{equation}
\delta ^{(n)}(s-\Phi )=\frac{d^{n}}{ds^{n}}\delta (s-\Phi ).
\end{equation}

For the interpolating current $J_{\mu }^{2}$ we find%
\begin{eqnarray}
&&\Pi _{\mu \nu }^{\mathrm{QCD}}(p,q)=i\int d^{4}xe^{ipx}\left\{ \left[
\gamma _{\nu }\widetilde{S}_{c}^{ib}(x)\gamma _{\mu }\widetilde{S}%
_{c}^{ai}(-x){}\gamma _{5}\right. \right.  \notag \\
&&\left. -\gamma _{5}\widetilde{S}_{c}^{ib}(x){}\gamma _{\mu }\widetilde{S}%
_{c}^{ai}(-x){}\gamma _{\nu }\right] _{\alpha \beta }\langle \phi (q)|%
\overline{s}_{\alpha }^{a}s_{\beta }^{b}|0\rangle  \notag \\
&&+\left[ \gamma _{\nu }\widetilde{S}_{c}^{ib}(x)\gamma _{\mu }\widetilde{S}%
_{c}^{bi}(-x){}\gamma _{5}-\gamma _{5}\widetilde{S}_{c}^{ib}(x){}\gamma
_{\mu }\widetilde{S}_{c}^{bi}(-x){}\gamma _{\nu }\right] _{\alpha \beta }
\notag \\
&&\left. \times \langle \phi (q)|\overline{s}_{\alpha }^{a}s_{\beta
}^{a}|0\rangle \right\} ,
\end{eqnarray}
The corresponding spectral density is
\begin{equation}
\rho_{c}^{(2)\mathrm{QCD}}(s)=2\rho_{c}^{(1)\mathrm{QCD}}(s),
\end{equation}
where $\rho_{c}^{(1)\mathrm{QCD}}(s)$ is given by Eq.\ (\ref{eq:Sdensity}).
\begin{widetext}

\begin{figure}[h!]
\begin{center}
\includegraphics[totalheight=6cm,width=8cm]{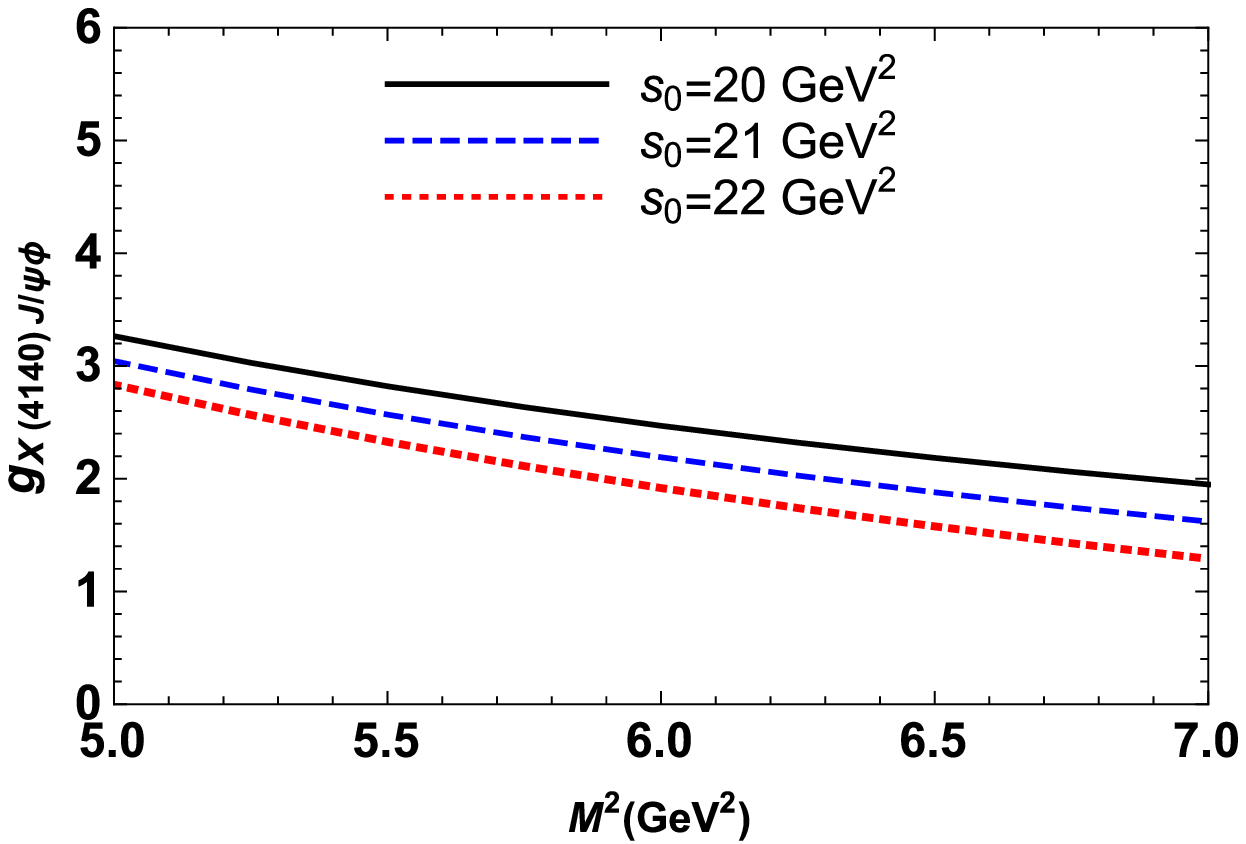}\,\, %
\includegraphics[totalheight=6cm,width=8cm]{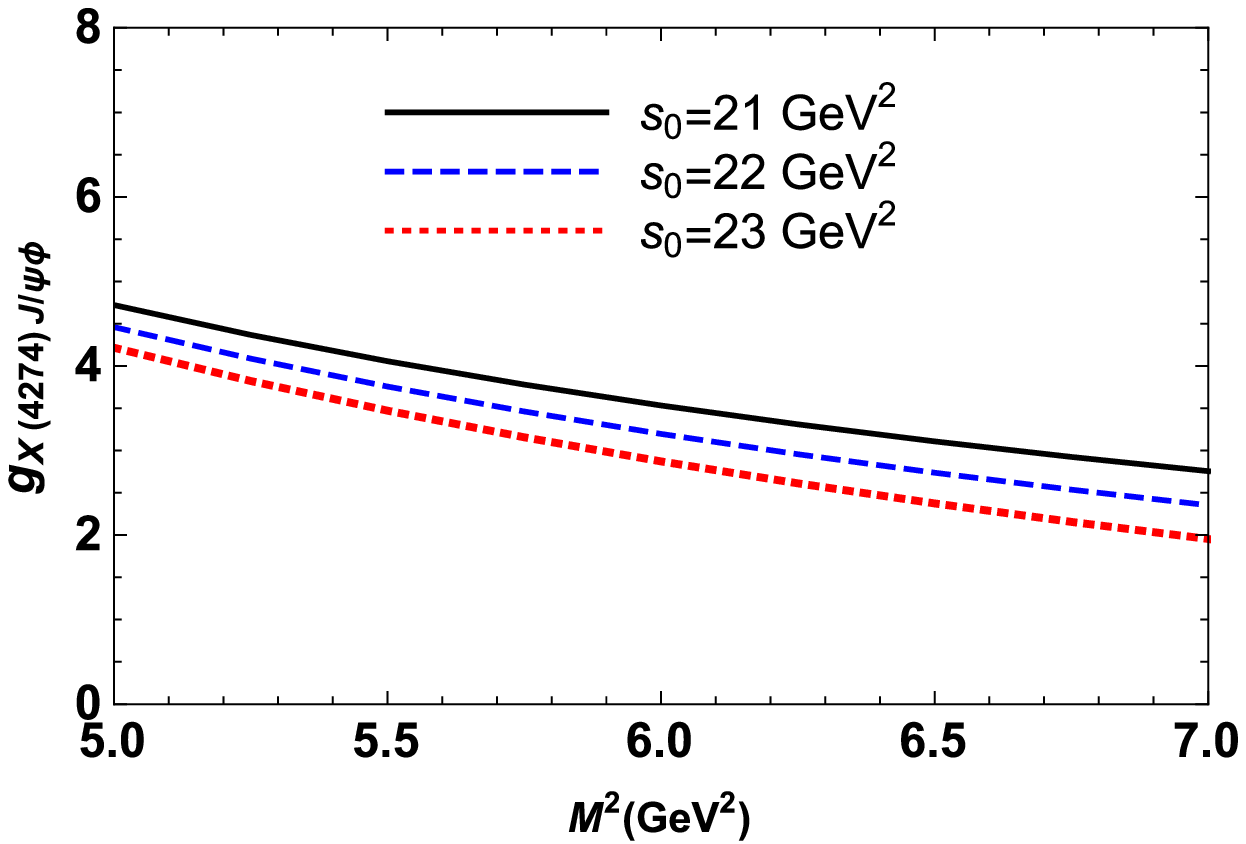}
\end{center}
\caption{ The strong coupling $g_{X_1J/\psi \phi}$ (left) and $g_{X_2J/\psi \phi}$
(right) as functions of the Borel parameter.}
\label{fig:VertexCoupl}
\end{figure}

\end{widetext}
The final expression for the strong coupling $g_{XJ/\psi \phi }$ has the
form
\begin{eqnarray}
g_{XJ/\psi \phi } &=&\frac{1}{f_{J/\psi }f_{X}m_{J/\psi }m_{X}}\left( 1-M^{2}%
\frac{d}{dM^{2}}\right) M^{2}  \notag \\
&&\times \int_{4m_{c}^{2}}^{s_{0}}dse^{(m^{2}-s)/M^{2}}\rho _{c}^{\mathrm{QCD%
}}(s).
\end{eqnarray}

The width of the decay $X\rightarrow J/\psi \phi $ is given by the formula%
\begin{eqnarray}
&&\Gamma (X\rightarrow J/\psi \phi )=\frac{\lambda (m_{X},m_{J/\psi
},m_{\phi })}{48\pi m_{X}^{4}m_{\phi }^{2}}g_{XJ/\psi \phi }^{2}\left[
\left( m_{X}^{2}+m_{\phi }^{2}\right) \right.  \notag \\
&& \times m_{J/\psi }^{4}+\left( m_{X}^{2}-m_{\phi }^{2}\right)
^{2}\left( m_{X}^{2}+m_{\phi }^{2}-2m_{J/\psi }^{2}\right)  \notag \\
&&\left.+4m_{X}^{2}m_{J/\psi}^{2}m_{\phi}^{2} \right],
\end{eqnarray}%
where $\lambda (a,b,c)$ is the standard function%
\begin{equation*}
\lambda (a,b,c)=\frac{\sqrt{%
a^{4}+b^{4}+c^{4}-2(a^{2}b^{2}+a^{2}c^{2}+b^{2}c^{2})}}{2a}.
\end{equation*}

The results of the numerical computations for the strong couplings and decay widths
are collected in Table\ \ref{tab:Results2A}. Here we also show  the working ranges for
the parameters $M^2$ and $s_0$, where the predictions for the couplings $g_{X_1J/\psi \phi}$
and $g_{X_2J/\psi \phi}$ are obtained. Within these ranges the sum rules
satisfy all requirements typical for such kind of calculations. Indeed, the  pole contribution
to the sum rule on the average amounts to $\sim 44\% $ of the result. The convergence of
OPE is fulfilled, too. Thus dimension-8 contribution constitutes only  $1\% $
of the sum rule.

In Fig.\ \ref{fig:VertexCoupl} we plot the couplings $g_{X_1J/\psi \phi}$
and $g_{X_2J/\psi \phi}$  as functions of the Borel parameter at fixed $s_0$. One can see
that the couplings are sensitive to the choice of the auxiliary parameters $M^2$ and $s_0$.
This sensitivity is a main source of theoretical ambiguities of the performed analysis,
numerical estimates of which can be found in Table\ \ref{tab:Results2A}, as well.

Comparing theoretical predictions and LHCb data, one sees that width of the decay
$X(4140) \to J/\psi \phi$ is in accord with the experimental data, whereas
$\Gamma(X(4274) \to J/\psi \phi)$ considerably exceeds and does not explain them.

\section{Discussion and concluding remarks}
\label{sec:Conclusions}
In the present work we have calculated the masses of the resonances
$X(4140)$ and $X(4274)$, and width of the decay channels $X(4140) \to J/\psi \phi$ and
$X(4274) \to J/\psi \phi$.
We have treated these resonances as the $1^{++}$  states in the multiplet of the color triplet
and sextet diquark-antidiquarks, respectively.
As is seen from Table\ \ref{tab:Results3A}, our predictions for the masses of $X(4140)$  and $X(4274)$,  obtained
using the two-point QCD sum rule method, are in a nice agreement with recent  measurements of the LHCb
Collaboration \cite{Aaij:2016iza}.

The  $X(4140)$  and $X(4274)$ states were  previously studied in Refs.\ \cite{Albuquerque:2009ak,Chen:2010ze,Chen:2016oma,Wang:2016tzr,Wang:2016dcb}.
Thus, the resonance $X(4140)$ was treated in Ref.\ \cite{Albuquerque:2009ak} as a molecule-like
bound state with $J^{PC}=0^{++}$ built of the mesons $D_s^{*}\bar{D}_s^{*}$.
Calculation of its mass, performed there using two-point QCD sum rule method and relevant
interpolating current gives a result, which correctly describes the experimental data. Nevertheless,
the LHCb Collaboration have excluded interpretation of the $X(4140)$ resonance as a molecule-like state.

As we have noted above, the masses of the $X(4140)$  and $X(4274)$ resonances
in the context of the two-point sum rule method were computed  also in Ref.\ \cite{Chen:2010ze}.
The obtained predictions  within errors explain the LHCb data  \cite{Chen:2016oma}.
Let us note that  $X(4140)$  and $X(4274)$ resonances were treated in
Refs.\ \cite{Chen:2010ze,Chen:2016oma} as the axial-vector states with
triplet and sextet color structures, respectively.

The investigations carried out in Ref.\ \cite{Wang:2016tzr} using sum rule approach and
two types of interpolating currents, however excluded interpretation of the $X(4140)$ resonance as a diquark-antidiquark state. The reason was that   $m_{X_1}$ extracted from the corresponding
sum rules either lay below LHCb data or overshot it (see, Table\ \ref{tab:Results3A}).

The $X(4274)$ was explored as
a molecule-like color octet state \cite{Wang:2016dcb}, and its mass $m_{X_2}$ was estimated as
\begin{equation}
 m_{X_2}=4.27 \pm 0.09\ \mathrm{GeV}.
\end{equation}
But  width of the decay $X(4274) \to J/\psi \phi$
\begin{equation}
\Gamma(X(4274) \to J/\psi \phi)=1.8\ \mathrm{GeV}
\end{equation}
evaluated in the framework of the three-point QCD sum rule approach,
considerably exceeded the LHCb value, therefore the author ruled
out the suggested interpretation of the $X(4274)$ state.

We have calculated the widths of  $X(4140/4274) \to J/\psi \phi$
decays, as well. The obtained predictions are collected in  Table\ \ref{tab:Results3A}.
It is evident, that our results for the  mass and width of the $X(4140)$ resonance
allow us to consider it as a serious candidate to the color triplet
$J^{PC}=1^{++}$ diquark-antidiquark state.
At the same time, interpretation of $X(4274)$ as a pure color sextet tetraquark which is,
in accordance with our results, a "wide" resonance, in the light of the LHCb data seems problematic: LHCb specifies it
as a narrow state. Perhaps  $X(4274)$
is an admixture of the color sextet tetraquark and a conventional charmonium. But this and alternative suggestions on the nature of the $X(4274)$ resonance require further investigations.

\begin{table}[tbp]
\begin{tabular}{|c|c|c|c|c|}
\hline\hline
 & $m_{X_1}$ & $\Gamma_{X_1}$ & $m_{X_2}$ & $\Gamma_{X_2}$  \\
 & $(\mathrm{MeV})$ & $(\mathrm{MeV})$ & $(\mathrm{MeV})$ & $(\mathrm{MeV})$  \\
 \hline\hline
LHCb& $4146\pm 4.5_{-2.8}^{+4.6}$ & $83\pm
21_{-14}^{+21}$ & $4273\pm 8.3_{-3.6}^{+17.2}$ & $56\pm
11_{-11}^{+8}$ \\ \hline
Our w. & $4183 \pm 115$ & $80 \pm 29$ & $4264 \pm 117$& $272 \pm 81$ \\ \hline
\cite{Albuquerque:2009ak}& $4140 \pm 90$ & $-$ &   $-$& $-$\\  \hline
\cite{Chen:2010ze}& $4070 \pm 100$ & $-$ &   $4220 \pm 100$& $-$\\ \hline
\cite{Wang:2016tzr}& $3950 \pm 90$ & $-$ &   $-$& $-$\\
& $5000 \pm 100$ & $-$ &   $-$& $-$\\ \hline
\cite{Wang:2016dcb}& $-$ & $-$ &   $4270 \pm 90$& $1800$\\  \hline
\hline\hline
\end{tabular}%
\caption{The LHCb data and theoretical predictions for the mass and decay width of the resonances $X(4140)$
and $X(4274)$.}
\label{tab:Results3A}
\end{table}

\section*{ACKNOWLEDGEMENTS}

K. A.  thanks  T\"{U}B\.{I}TAK for partial financial support provided under the grant no: 115F183.

\appendix*

\section{ The $s$ and $c$-quark propagators}

\renewcommand{\theequation}{\Alph{section}.\arabic{equation}} \label{sec:App}

The light and heavy quark propagators are the important quantities to find
QCD side of the correlation functions in both the mass and strong coupling
calculations. We employ the $s$- quark propagator $S_{s}^{ab}(x)$, which is
given by the following formula%
\begin{eqnarray}
&&S_{s}^{ab}(x)=i\delta _{ab}\frac{\slashed x}{2\pi ^{2}x^{4}}-\delta _{ab}%
\frac{m_{s}}{4\pi ^{2}x^{2}}-\delta _{ab}\frac{\langle \overline{s}s\rangle
}{12}  \notag \\
&&+i\delta _{ab}\frac{\slashed xm_{s}\langle \overline{s}s\rangle }{48}%
-\delta _{ab}\frac{x^{2}}{192}\langle \overline{s}g_{s}\sigma Gs\rangle
+i\delta _{ab}\frac{x^{2}\slashed xm_{s}}{1152}  \notag \\
&&\times \langle \overline{s}g_{s}\sigma Gs\rangle -i\frac{%
g_{s}G_{ab}^{\alpha \beta }}{32\pi ^{2}x^{2}}\left[ \slashed x{\sigma
_{\alpha \beta }+\sigma _{\alpha \beta }}\slashed x\right]  \notag \\
&&-i\delta _{ab}\frac{x^{2}\slashed xg_{s}^{2}\langle \overline{s}s\rangle
^{2}}{7776}-\delta _{ab}\frac{x^{4}\langle \overline{s}s\rangle \langle
g_{s}^{2}G^{2}\rangle }{27648}+\ldots  \label{eq:qprop}
\end{eqnarray}%
For the $c$-quark propagator $S_{c}^{ab}(x)$ we employ the well-known
expression
\begin{eqnarray}
&&S_{c}^{ab}(x)=i\int \frac{d^{4}k}{(2\pi )^{4}}e^{-ikx}\Bigg \{\frac{\delta
_{ab}\left( {\slashed k}+m_{c}\right) }{k^{2}-m_{c}^{2}}  \notag \\
&&-\frac{g_{s}G_{ab}^{\alpha \beta }}{4}\frac{\sigma _{\alpha \beta }\left( {%
\slashed k}+m_{c}\right) +\left( {\slashed k}+m_{c}\right) \sigma _{\alpha
\beta }}{(k^{2}-m_{c}^{2})^{2}}  \notag \\
&&+\frac{g_{s}^{2}G^{2}}{12}\delta _{ab}m_{c}\frac{k^{2}+m_{c}{\slashed k}}{%
(k^{2}-m_{c}^{2})^{4}}+\frac{g_{s}^{3}G^{3}}{48}\delta _{ab}\frac{\left( {%
\slashed k}+m_{c}\right) }{(k^{2}-m_{c}^{2})^{6}}  \notag \\
&&\times \left[ {\slashed k}\left( k^{2}-3m_{c}^{2}\right) +2m_{c}\left(
2k^{2}-m_{c}^{2}\right) \right] \left( {\slashed k}+m_{c}\right) +\ldots %
\Bigg \}.  \notag \\
&&{}  \label{eq:Qprop}
\end{eqnarray}%
In Eqs.\ (\ref{eq:qprop}) and (\ref{eq:Qprop}) we adopt the notations
\begin{eqnarray}
&&G_{ab}^{\alpha \beta }=G_{A}^{\alpha \beta
}t_{ab}^{A},\,\,~~G^{2}=G_{\alpha \beta }^{A}G_{\alpha \beta }^{A},  \notag
\\
&&G^{3}=\,\,f^{ABC}G_{\mu \nu }^{A}G_{\nu \delta }^{B}G_{\delta \mu }^{C},
\label{kabi}
\end{eqnarray}%
with $a,\,b=1,2,3$ being the color indices, and $A,B,C=1,\,2\,\ldots 8$ \ .
In Eq. (\ref{kabi}),  $t^{A}=\lambda ^{A}/2$, $\lambda ^{A}$ are the Gell-Mann matrices,
and the gluon field strength tensor $G_{\alpha \beta }^{A}\equiv G_{\alpha
\beta }^{A}(0)$ is fixed at $x=0$.

\end{document}